\documentclass[11pt]{article}
\usepackage[all]{xy}
\xyoption{arc}
\xyoption{curve}
\usepackage{color}
\usepackage[numbers]{natbib}
\usepackage{float}

\usepackage{geometry} 
\usepackage[applemac]{inputenc} 
\usepackage{textcomp} 
\usepackage{graphicx}  
\usepackage{amsmath,amssymb}  
\usepackage{bm}  

\usepackage{amsfonts}

 \usepackage[margin=10pt, font=small, labelfont=bf]{caption}

\def\inf{\mathop{\hbox{inf}}}

\newtheorem{teor}{Theorem}[section]

\newtheorem{remar}[teor]{Remark}

\newcommand{\fdim}{\hspace*{\fill}$\Box$}
\newcommand{\dimostr}{{\bf Proof: }}

\newcommand{\comment}[1]{}

\newcommand*{\rom}[1]{\expandafter\@slowromancap\romannumeral #1@}
\newcommand{\RNum}[1]{\uppercase\expandafter{\romannumeral #1\relax}}

\begin{document}
\title{A note on local uniqueness of equilibria: How isolated is a local equilibrium?
}
\author{Stefano Matta\thanks{Correspondence to
S. Matta, Dipartimento di Economia, University of Cagliari, 
viale S. Ignazio 17, 09123 Cagliari, Italy,
tel. 00390706753340, Fax  0039070660929
E-mail: smatta@unica.it.
The author was supported 
by STAGE, funded by Fondazione di Sardegna and Regione Autonoma della Sardegna.
}. 
\\
\and\centerline{(University of Cagliari)}}

\maketitle

\vspace{0.4in}

\noindent\textbf{Abstract}:
The motivation of this note is to show how singular values 
affect local uniqueness.
More precisely, Theorem 3.1 shows how to construct 
a neighborhood (a ball) of a regular equilibrium whose diameter
represents an estimate of local uniqueness, hence 
providing a measure of how isolated
a (local) unique equilibrium can be. The result, whose relevance 
in terms of comparative statics is evident, is based on reasonable 
and natural assumptions and hence is applicable in many different settings,
ranging from pure exchange economies to non-cooperative games.

\vspace{0.3in}

\vspace{0.3in}

  \newpage

\section{Introduction}\label{sec_intro}

Consider an equilibrium
equation $f(p,q)=0$,  where $p$ and $q$ denote the unknowns and parameters, respectively.
Local uniqueness and continuous (smooth) dependence of the unknowns on the parameters
are at the heart of comparative statics.
This means that, roughly speaking, the (locally unique) solution $p$ changes continuously (smoothly)
as the parameter $q$ varies under a continuous (smooth) perturbation.

More precisely, at a point $p$ belonging to the solution set of $q$,
there exists a neighborhood $N$ of $(p,q)$ such that the projection onto the second factor, 
$pr:(p,q)\to q$, restricted to $N$, is an homeomorphism (diffeomorphism).

This regularity property has been extensively studied in the literature in different settings,
in particular after the introduction of the differentiable viewpoint in the seminal paper by \cite{debreu}.
The study of the equilibrium set $E=\{(p,q)|\,f(p,q)=0\}$ and the properties of
the projection $pr:E\to Q$, $(p,q)\mapsto q$, allows to establish natural connections
between economic and mathematical properties:
e.g., the surjectivity of $pr$ (existence), the cardinality of the set $\{pr^{-1}(q)\}$ 
(uniqueness/multiplicity), the continuity of the correspondence 
of $pr^{-1}$ (structural stability).

Under the very restrictive hypothesis of global uniqueness, i.e. when the set $pr^{-1}(q)$ is a single-tone set for every parameter $q$, 
the whole solution set $E$ is homeomorphic (diffeomorphic) to the parameter set.

Multiplicity, on the other hand, is deeply related to the set of singular values of the projection.
Even if this set has measure zero in $Q$ under certain assumptions,
its components of codimension-one and codimension-two  can be relevant.
Consider, for example, redistribution policies, represented by
continuous perturbations of the parameters. In such an instance,
it is crucial not to cross the singular set to avoid catastrophes and 
undesiderable welfare effects, that can be determined by prices multiplicity.
For example, Figure 1, taken from \citep{die},
shows an example of two policies sharing the same target, but
whose (different) outcomes depend on the order of use of resources.

\begin{figure}[h]\label{figdie}
\centering
\includegraphics[scale=.25]{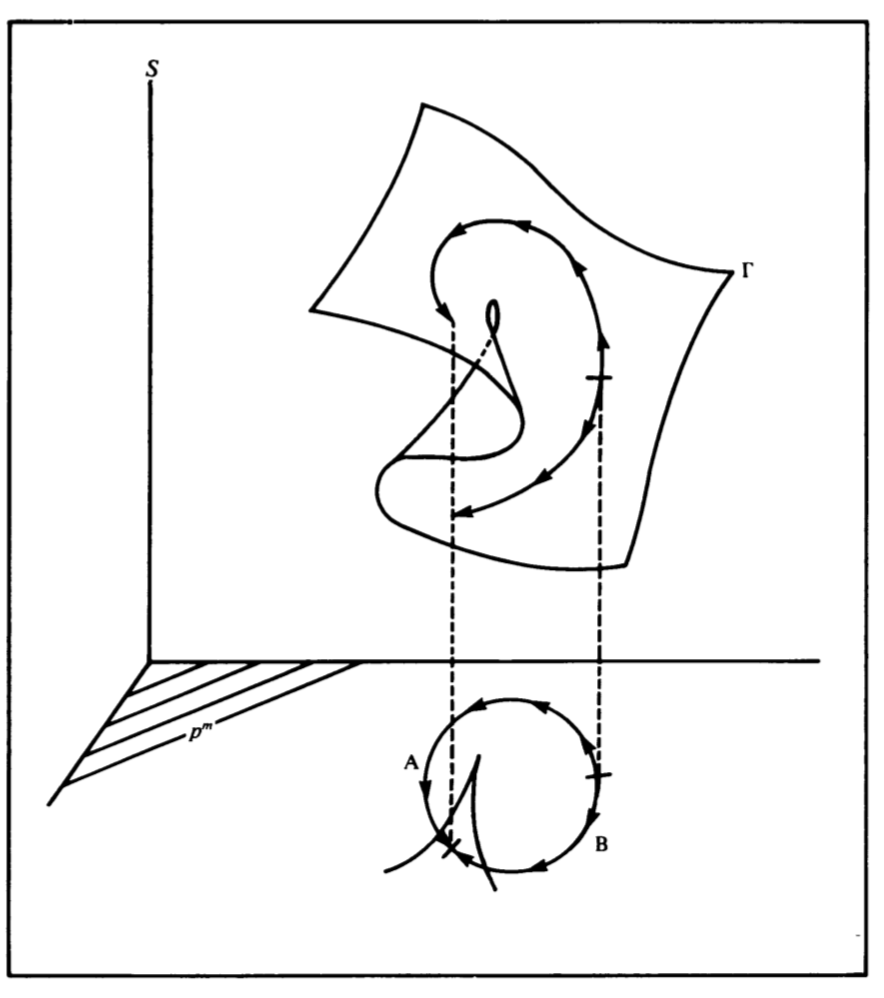}
\caption{Two policies sharing the same target but with different outcomes (taken from \citep{die}).}
\end{figure}

The codimension-two components also play a key role, as highlighted in this note,
since they have an impact on the topology of the
parameter set. The intuition behind this is that they can be seen 
as holes inside $Q$. 
The consequences of this property have not, to the best of my knowledge, 
been analyzed in the literature, which has traditionally focused on the codimension-one component,
the only one that can disconnect $Q$.

The motivation of this short note is to highlight how the singular values can affect the size of local uniqueness.
This information can be used to estimate how isolated a (local) unique equilibrium can be.
More precisely, Theorem \ref{teoruniq} shows a sufficient condition that enables to construct a neighborhood $N$ such that 
the projection restricted to $N$ is injective. This means that local uniqueness is satisfied in $N$.

This result can be applied to different settings,
characterized by multiplicity and multiple roots associated to critical solutions.
For example, one can appraise potential distorsive effects (e.g. transfer paradox \cite{batra}) 
of redistributive policies  \cite{gago} in exchange and production economies \cite{balib2}.
This is due to the fact that a common geometric structure is present. The same structure, with 
suitable changes, can be found, e.g.,  in the case of infinite economies  \cite{bainf,chiz} or non-cooperative game theory \cite{gw}.

This paper is organized as follow. Section \ref{sec_def} recalls the main definitions
and tools. Section \ref{sec_main} is devoted to the proof of the main result.


\section{Definitions}\label{sec_def}

In this section the main definitions and properties are outlined for the reader's convenience.
For an introduction to covering spaces, the reader is referred
to \cite{massey}.

The analysis is focused on the projection map $pr:E\to Q$. The set
$Q$ denotes the parameter set and $E$ is the equilibrium set, i.e.
$E=\{(p,q)|\,f(p,q)=0\}$, where $f(p,q)=0$ represents an equilibrium condition
to be satisfied by the unknowns $p$, given the constraints and the primitives of the model.

Suppose that  both $E$ and $Q$ are smooth   manifolds where we can measure  the length of curves and the distance between points.
More precisely, given a  piecewise differentiable curve $\tilde \gamma:[0, 1]\rightarrow E$  joining two points $\tilde x_1$ and $\tilde x_2$ in $E$, namely $\tilde\gamma (0)=\tilde x_1$ and $\tilde\gamma (1)=\tilde x_2$,
we can compute the length of $\tilde \gamma$ denoted by  $L_E(\tilde \gamma)$  and the distance
\begin{equation}\label{lengdef}
d_E(\tilde x_1, \tilde x_2)=\inf_{\tilde \gamma\in\tilde\Gamma_{\tilde x_1, \tilde x_2}}L_E(\tilde\gamma).
\end{equation}
between $\tilde x_1$ and $\tilde x_2$,  
where $\tilde \Gamma_{\tilde x_1, \tilde x_2}$ denotes the set of  piecewise differentiable curves
 with endpoints $\tilde x_1$ and $\tilde x_2$.
Similarly, we can compute the length $L_Q(\gamma)$ of    a  piecewise differentiable curve $\gamma:[0, 1]\rightarrow Q$  joining two points $x_1$ and $x_2$ of $Q$
and their distance $d_Q(x_1, x_2)$.

In this paper we assume that the following properties are satisfied:
\begin{itemize}
\item [(i)]
$pr$ is smooth;
\item [(ii)]
$pr$ is proper;
\item [(iii)]
$pr$ is a length  decreasing map, i.e. 
\begin{equation}\label{Ldec}
L_E(\tilde\gamma)\geq L_Q(pr\circ\tilde\gamma), \ \forall\tilde\gamma:[0, 1]\to E.
\end{equation}
Consequently, by \eqref{lengdef},  $pr$ is distance decreasing, namely
\begin{equation}\label{ddec}
d_E(\tilde x, \tilde y)\geq d_Q(pr(\tilde x), pr(\tilde y)),\ \forall \tilde x, \tilde y\in E.
\end{equation}
\item [(iv)]
the ball centered at  a point $\tilde x\in E$ (resp. $x\in Q$) of radius $\tilde r$ (resp. $r$), namely 
$B_{\tilde r}(\tilde x)=\{\tilde y\in E \ | \ d_E(\tilde y, \tilde x)<\tilde r\}$ (resp. $B_{r}(x)=\{y\in Q \ | \ d_Q( y,  x)< r\}$ ) is connected.
\end{itemize}
All these are reasonable and natural  assumptions.
Smoothness is an approximation of continuity under suitable topologies,  properness
also has a nice economic meaning:  for example,
it represents the idea of scarcity and desirability. 
Assumption (iii) and (iv) have a very intuitive meaning. Indeed it is expected that 
a projection does not increase the length of curves and that
the metrics chosen satisfy the connectedness property.  
As a specific and natural example, one can take a Riemannian metric $g_Q$ on $Q$ 
(if $Q$ is a subset of some Euclidean space, $g_Q$ can be the flat metric
as is the case, e.g.,  in literature related to the Edgeworth box)
and the pull-back metric $g_E=pr^*g_Q$ on $E$. 
Then $pr$ is length decreasing if we endow $Q$ and $E$ with the 
length functions associated to these metrics.

In this setting, the application of the inverse function theorem (IFT) and Sard's theorem
to $pr$ leads to standard regularity properties enjoyed by an open and dense subset of the parameter space, denoted by $\cal R$. More precisely, 
it is not hard to prove
that  $pr_{|pr^{-1}(\cal R)}:pr^{-1}(\cal R)\rightarrow {\cal R}$ is a covering map (see \cite[Proposition 2.2]{lmstr}). Indeed  the  properness of $pr$ is a sufficient condition to turn a surjective local diffeomorphism
into a covering map.  
We recall that a {\em covering map} between two topological spaces $\tilde X$ and $X$ is a continuous surjective 
map such that each $x\in X$ admits a {\em well-covered} neighborhood, i.e.,
each $x\in X$ has an open neighborhood $U$ such that $p^{-1}(U)$ is a disjoint union of open sets in $\tilde X$, each of which is mapped by $p$ homeomorphically onto $U$.

The economic meaning behind the covering property is that it represents  
the well-known property of smooth selection of equilibria.
This  is crucial for comparative statics analysis 
and characterizes 
the regular values ${\cal R}$ of the projection. 
Its complement $Q\setminus \cal R$
represents the projection of the critical equilibria,
denoted by $\Sigma=\cup \Sigma^k$. It is the union of closed sets of
codimension  $k$, with $k$ greater or equal to $1$. Hence $\Sigma^1$ denotes the 
component which disconnects the parameter set. For a deep analysis
of the set of singular and critical equilibra in a
general equilibrium framework, the reader is referred to
\cite{basing, bacrit, balib2,lmcrit}.
Literature is usually focused on
$\Sigma^1$ as the main cause of catastrophes.
But $\Sigma^2$ affects ${\cal R}$'s topology in a relevant way as we will see 
in Theorem \ref{teoruniq} (see also Remark \ref{remtheor} below).

\comment{
I end this section recalling the {\em arc lifting property} \citep[Lemma 3.1]{massey},
that will be used in the proof of our main result.
Given a smooth map $p:\tilde X\to X$, a {\em lift} 
of a map $f : Y \to X$, where $Y$ is a topological space, 
is a map $\tilde f : Y \to \tilde X$ such that $p\tilde f = f$.
Let $\alpha: [0,1] \to X$ be an arc. If $p$ is a covering map, then
a unique lift of $\alpha$,
$\tilde\alpha: [0,1] \to \tilde X$, exists, whose starting point $\tilde x_0$ 
is any point chosen in the fiber of $p^{-1}(\alpha(0))$.
}


\section{Main result}\label{sec_main}

Let  $x\in {\cal R}$ and let $\Gamma_x$ be the 
set of  non contractible loops $\gamma: [0, 1]\to {\cal R}$,
$\gamma(0)=\gamma(1)=x$. 
Notice that the image of $\gamma$ lies in a connected component $K_x$ of ${\cal R}$ containing $x$.
Let  
\begin{equation}\label{mx}
m_x=\inf_{\gamma\in \Gamma_x}L_{d_Q}(\gamma)\geq 0
\end{equation}
be the nonnegative  number which measures the length of the minimal loop of all the non contractible closed curves in $K_x$ based
at $x$. If no $\Sigma^2$ component belongs to the connected component of $Q\setminus \Sigma^1$ containing $x$, then $m_x=0$.
It could be said in a suggestive way that $\Sigma^2$ are  \lq\lq holes'' in $K_x$ iff  $m_x\neq 0$.

Let us also  define 
\begin{equation}\label{dx}
d_x=d_{Q}(x,\Sigma^1)=\inf_{\sigma\in \Sigma^1}d_{Q}(x,\sigma).
\end{equation}
The positive number
$d_x$ measures the distance between $x$
and the boundary of $K_x$, which corresponds to $\Sigma^1$.
This distance is well-defined since $\Sigma^1$
is closed in $Q$, being the projection of a closed set via a proper map.

The following theorem uses the information so far
to construct a  neighborhood $N$ of $\tilde x$ such that $pr_{|N}$ is injective,
i.e. where one has  local uniqueness.

\begin{teor}\label{teoruniq}
Let us assume assumptions (i)-(iv) above are satisfied.
Let $x\in {\cal R}$ and $\tilde x\in pr^{-1}(x)$. 
Let  $r_x$ be the positive real number defined as

$$r_x=\begin{cases}
 3d_x &  \mbox{if}\ m_x= 0\\
 \min(m_{x}, d_{x}) &\mbox{if}\ m_x\neq 0.
\end{cases}$$
where $m_x$ and $d_x$ are given by \eqref{mx} and \eqref{dx} respectively. 
Then $pr_{|_{B_\frac{r_x}{3}(\tilde x)}}$ is injective.

\end{teor}

\noindent \dimostr\rm 
Denote by $\tilde K_x$
the connected component of $pr^{-1}({\cal R})$ containing $\tilde x$.
Then, being $E$  and $Q$ smooth manifolds, by \cite[Ch. 5, Lemma 2.1]{massey} $pr_{|\tilde K_x}:\tilde K_x\to K_x$
is a covering map. If $m_x=0$ then $K_x$ is simply-connected and hence  $pr_{|\tilde K_x}$ \cite[Ch. 5, Exercise 6.1]{massey} is a diffeomorphism.
Notice that by (iv) $B_{d_x}(x)$ is connected and hence $B_{d_x}(x)\subset K_x$. Moreover since by \eqref{ddec} $pr$ is distance decreasing 
one gets that $pr\left(B_{d_x}(\tilde x)\right)\subset B_{d_x}(x)\subset K_x$ and hence $B_{d_x}(\tilde x)\subset \tilde K_x$. 
It follows that  $pr_{|_{B_\frac{r_x}{3}(\tilde x)}}=pr_{|B_{d_x}(\tilde x)}$ is injective.

Let $m_x\neq 0$. Assume,  by contradiction, that $\tilde x_1$ and $\tilde x_2$ are two distinct points in $B_\frac{r_x}{3}(\tilde x)$, $r_x=\min(m_{x}, d_{x})$,
such that $pr(\tilde x_1)=pr(\tilde x_2)=x$.  
Let $\tilde \Gamma_{\tilde x_1, \tilde x_2}$ denote the set of  piecewise differentiable curves $\tilde \gamma :[0, 1]\to B_\frac{r_x}{3}(\tilde x)$
 with endpoints $\tilde x_1$ and $\tilde x_2$.
 Notice that $\tilde \Gamma_{\tilde x_1, \tilde x_2}\neq\emptyset$ by assumption (iv). 
Then, for each $\tilde\gamma\in \tilde \Gamma_{\tilde x_1, \tilde x_2}$, the curve 
$$\gamma =pr\circ\tilde\gamma:[0,1]\to K_x$$ is a piecewise 
differentiable loop based on $x$.
By \cite[Ch. 5, Theorem 4.1]{massey} each of these  loops  is non-contractible and hence by \eqref{mx} one has  $L_{d_Q}(\gamma)\geq m_x$.
It follows, by \eqref{lengdef} and \eqref{Ldec},
that 
$$d_E(\tilde x_1, \tilde x_2)=\inf _{\tilde\gamma\in \tilde\Gamma_{\tilde x_1, \tilde x_2}}L_{d_E}(\tilde \gamma)\geq\inf _{\gamma\in \Gamma_{x}} L_{d_Q}(\gamma)=m_x\geq r_x.$$
On the other hand, by the triangle inequality,
$$d_{E}(\tilde x_1,\tilde x_2)\leq d_{E}(\tilde x_1,\tilde x)+ d_{E}(\tilde x,\tilde x_2)=\frac{r_x}{3}+\frac{r_x}{3}=
\frac{2}{3}r_{x} <  r_{x},$$
yielding the desired contradiction.
\fdim

\begin{remar}\label{remtheor}\rm 
The key idea behind the proof of the theorem is, 
roughly speaking, as follows. If $K_x$ is the  connected component of ${\cal R}$ containing $x$,
the set $\Sigma^2$ are \lq\lq holes'' inside $K_x$.
If $K_x$ did not contain holes, then it would be
simply connected and hence the solution set would be single-tone (global uniqueness)  for every parameter in $K_x$,
if one restricts $pr$ to the  connected component $\tilde K_x$ of $pr^{-1}({\cal R})$ containing $\tilde x$. 
On the other hand, if $K_x$ contains holes then $m_x\neq 0$, namely 
one has to be careful with the non-contractible loops by  taking the ball centered at  $\tilde x$ of radius $\min(d_{x}, m_{x})$.
\end{remar}

\begin{remar}\label{remtheor}\rm 
The diameter of the ball $\frac{r_x}{3}(\tilde x)$ 
represents a sufficient condition that ensures
injectivity but it is does not maximize the size of the ball (if  $m_x\neq 0$).
It can be seen as an estimate of local uniqueness
and a measure  of how isolated is a local equilibrium, providing us with a possible answer to the question raised in the title the paper.
 \end{remar}





\small{}

\end{document}